\documentclass[oribibl]{llncs}
  
\usepackage[active]{srcltx} 
\usepackage[centertags]{amsmath} 
\usepackage{stackrel}
\usepackage{amssymb}
\usepackage{graphicx}
\usepackage{extpfeil}
\usepackage{ulem}
\usepackage{textcase}
\usepackage[ruled,vlined]{algorithm2e}
\usepackage{color}
\usepackage{graphics}
\usepackage{makeidx}

\DeclareMathOperator*{\argmax}{arg\,max}

\makeatletter
\newif\if@restonecol
\makeatother

\begin{document}
\normalem

\title{GPU-Based Heuristic Solver for Linear Sum Assignment Problems Under Real-time Constraints}

\author{Roberto Roverso\inst{1,2} \and Amgad Naiem\inst{1,3} \and
  Mohammed El-Beltagy\inst{1,3} \and  Sameh El-Ansary\inst{1,4}}
\institute{Peerialism Inc., Sweden
\and
KTH-Royal Institute of Technology, Sweden
\and
Cairo University, Egypt
\and
Nile University, Egypt\\
\email{\{roberto,amgad,mohammed,sameh\}@peerialism.com}\\
}

\date{}
\maketitle

\begin{abstract}
In this paper we modify a fast heuristic solver for the  Linear Sum Assignment
Problem (LSAP) for use on Graphical Processing Units (GPUs). The motivating scenario is an
industrial application for P2P live streaming that is moderated by a
central node which is periodically solving LSAP instances for assigning peers to one another. The central node needs to handle LSAP instances involving thousands of peers in as near to real-time as possible. Our findings are generic enough to be
applied in other contexts. Our main result is a parallel version of a
heuristic algorithm called Deep Greedy Switching (DGS) on GPUs using
the CUDA programming language. DGS sacrifices absolute optimality in
favor of low computation time and was designed as an alternative to
classical LSAP solvers such as the Hungarian and auctioning methods. The
contribution of the paper is threefold: First, we present the process
of trial-–and-–error we went through, in the hope that our experience
will be beneficial to adopters of GPU programming for similar
problems.  Second, we show the modifications needed to parallelize the
DGS algorithm. Third, we show the performance gains of our approach
compared to both a sequential CPU-based implementation of DGS and a
parallel GPU-based implementation of the auctioning algorithm.
\end{abstract}


\section{Introduction}
In order to deal with hard optimization or combinatorial problems in
time-constrained environments, it is often necessary to
sacrifice optimality in order to meet the imposed deadlines. In our work, we have dealt
with a large scale peer-to-peer live-streaming platform where the
task of assigning $n$ providers to $m$ receivers is carried out by a
centralized optimization engine. 
The problem of assigning peers to
one-another is modelled as a \emph{linear sum assignment problem} (LSAP). 
However, in our p2p system, the computational overhead
of minimizing the cost of assigning $n$ jobs (receivers) to $n$
agents (senders)  is usually 
quite high because we are often dealing with tens of thousands of agents
and jobs (peers in the system). We have seen our implementation of classical LSAP solvers take several hours to provide an optimal solution to a problem of this magnitude. 
In the context of live streaming we could only afford a few seconds to carry out this optimization. It was also important for us not to sacrifice optimality too much in the pursuit of a practical optimization solution. We hence opted for a strategy of trying to discover a fast heuristic near-optimal solver for LSAP that is also amenable to parallelization in such a way that can exploit the massive computational potential of modern GPUs. 
After structured experimentation on a number of ideas for a heuristic
optimizer, we found a simple and effective heuristic we called Deep
Greedy Switching~\cite{greedy} (DGS). It was shown to work extremely
well on the instances of LSAP we were interested in, and we never
observed it deviate from the optimal solution by more than $0.6$\%,
(c.f. \cite[p. 5]{greedy}). Seeing that DGS has parallelization
potential, we modified and adapted it to be run on any parallel
architecture and consequently also on GPUs.

In this work, we chose CUDA~\cite{cuda} to be our choice as a GPU
programming language to
implement the DGS solver. CUDA is a sufficiently general C-like language which allows for execution
of any kind of user-defined algorithms on the highly parallel architecture
of NVIDIA GPUs.
GPU programming has become increasingly popular in the
scientific community during the last few years. However, the task of developing whatsoever mathematical process in a
GPU-specific language still involves a fair amount of effort in understanding the 
hardware architecture of the target platform. In addition, implementation efforts
must take into consideration a set of best practices to achieve
best performance. 
This is the reason why in this paper we will provide an introduction to CUDA in
Section~\ref{sec:cuda}, in order to better understand its advantages,
best practices and limitations, so that it will later be easier to
appreciate the solver implementation's design choices in Section~\ref{sec:dgscuda}.
We will also detail the inner functioning of the DGS heuristic in
Section~\ref{sec:dgs} and
show results obtained by comparing both different versions of the DGS
solver and the final version of the GPU DGS solver with an implementation of the auction
algorithm running on GPUs in Section~\ref{sec:meas}. We will conclude the paper
with few considerations on the achievements of this work and our future plans in Section~\ref{sec:conclusion}.

\section{GPUs and the CUDA language}\label{sec:cuda}
Graphical Processing Units are mainly accelerators for graphical
applications, such as games and 3D modelling software, which make use
of the OpenGL and DirectX programming interfaces. Given that many of calculations involved in those applications are amenable to parallelization, GPUs have hence been architected as
massive parallel machines. In the last few years GPUs have
ceased to be exclusively fixed-function devices and have evolved to become flexible
parallel processors accessible through programming
languages~\cite{cuda}\cite{opencl}.  In fact, modern GPUs as NVIDIA's Tesla and GTX are
fundamentally fully programmable many-core chips, each one of them 
having a large number of parallel processors. Multicore chips are
called Streaming Multiprocessors (SMs) and their number can vary from
one, for low-end GPUs, to as many as thirty. Each SM contains in turn
8 Scalar Processors (SPs), each equipped with a set of
registers, and 16KB on-chip memory called Shared Memory. This
memory has lower access latency and higher bandwidth compared to off-chip memory, called Global
Memory, which is usually of the DDR3/DDR5 type and of size of
512MB to 4GB.

We chose CUDA as GPU Computing language for implementing our
solver because it best
accomplishes a trade-off between ease-of-use and required knowledge of the
hardware platform's architecture. Other GPU specific languages, such as AMD's
Stream~\cite{stream} and Kronos' OpenCL standard~\cite{opencl} look promising but fall short of
CUDA either for the lack of support and documentation or for the quality
of the development platform in terms of stability of the provided tools, such as
compilers and debuggers.
Even though  CUDA provides a sufficient degree of abstraction from the GPU
architecture to 
ease the task of implementing parallel algorithms, one must still understand the
basics of the functioning of NVIDIA GPUs to be able to fully utilize the
power of the language.

The CUDA programming model imposes the application to be organized
in a \emph{sequential part} running on a \emph{host}, usually the
machine's CPU, and parallel parts called \emph{kernels} that
execute code on a parallel \emph{device}, the GPU(s). Kernels
are
blocks of instructions which are executed across a number of parallel
threads. Those are then logically grouped by CUDA in a grid whose sub-parts are the thread blocks. The size of the grid and thread blocks are defined by the programmer.
This organization of threads derives from the legacy purpose of GPUs where the rendering of a texture (see grid) needs to be parallelized by assigning one thread to every pixel, which are then executed in batches (see thread blocks). A thread block is
a set of threads which can cooperate among themselves exclusively using barrier
synchronization, no other synchronization primitives are
provided. Each block has access to an amount of Shared Memory which is
exclusive for its group of threads to utilize. The blocks are therefore a way for CUDA to abstract the
physical architecture of Scalar Multiprocessors and Processors away from
the programmer. Management of Global and Shared Memory must be enforced
explicitly by the programmer through primitives provided by
CUDA. Although Global memory is sufficient to run any CUDA program, it
is advisable to use Shared Memory in order to obtain efficient
cooperation and communication between threads in a block. It is
particularly advantageous to let threads in a block load data from
global memory to shared on-chip memory, execute the kernel
instructions and later copy the result back in global memory.
\section{DGS Heuristic}
\label{sec:dgs}
In LSAP we try to attain the optimal assignment of $n$ agents to $n$ jobs, where there is a certain benefit $a_{ij}$ to be realized  when assigning agent $i$ to job $j$. The optimal assignment of agents to jobs is the one that yields the maximum total benefit, while respecting the constraint that each agent can only be assigned to only one job, and that no job is assigned to more than one agent. The assignment problem can be formally described as follows
\begin{align*}
 \max \sum_{i=1}^{n}&{\sum_{j=1}^{n}{a_{ij} x_{ij}}} \\
 \sum_{i=1}^{n}{x_{ij}} = 1 \quad & \quad \forall j \in \{1 \ldots n \} \\
\sum_{j=1}^{n}{x_{ij}} = 1 \quad & \quad \forall i \in \{1 \ldots n \} \\
 x_{ij} \in \{0,1\} \quad & \quad \forall i,j \in \{1 \ldots n \} 
\end{align*} 
There are many applications that involve LSAP, ranging from image processing to inventory management. The two most     popular algorithms for LSAP are the Hungarian method \cite{Kuhn1955} and the auction algorithm \cite{bertsekas1988aad}. The auction algorithm has been shown to be very effective in practice, for most instances of the assignment problem, and it is considered to be one of the fastest algorithms that guarantees a very near optimal solution (in the limit of $n\epsilon$). The algorithm works like a real auction where agents are bidding for jobs. Initially,  the price for each job is set to zeros and all agents are unassigned. At each iteration, unassigned agents bid simultaneously for their ``best'' jobs which causes the jobs' prices ($p_j$) to rise according. The prices work to diminish the net benefit ($a_{ij}-p_j$) an agent attains when being assigned a given job. Each job is awarded to the highest bidder, and the algorithm keeps iterating until all agents are assigned.

Although the auction algorithm is quite fast and can be easily parallelized,  it is not well suited to situations where  large instances of the assignment problem are involved and there is deadline after which a solution would be useless. Recently, a novel heuristic approach called \emph{Deep Greedy Switching (DGS)} \cite{greedy} was introduced for solving the assignment problem. It sacrifies very little in terms of optimality, for a huge gain in the running time of the algorithm over other methods. The DGS algorithm provides no guarantees\footnote{The authors are still working on a formal analysis of DGS that would help explain its surprising success. } for attaining an optimal solution, but in practice we have seen it deviate with less than 0.6\% from the optimal solutions, that are reported by the auction algorithm, at its worst performance. Such a minor sacrifice in optimality is acceptable in many dynamic systems where speed is the most important factor as an optimal solution that is delivered too late is practically useless. 
Compared with the auction algorithm, DGS has the added advantage that it starts out with a full assignment of jobs to agents and keeps improving that assignment during the course of its execution. The auction algorithm, however, attains full assignment only at termination. Hence, if a deadline has been reached where an assignment must be produced, DGS can interrupted to get the best assignment solution it has attained thus far. 
The DGS algorithm, shown in Algorithm~\ref{alg:dgs}, starts with a random initial solution, and then keeps on moving in a restricted 2-exchange neighborhood of this solution according to certain criteria until no further improvements are possible. 

\subsection{Initial Solution} The simplest way to obtain an initial solution is by randomly assigning jobs to agents.  An alternative is to do a greedy initial assignment where the benefit $a_{ij}$ is taken into account. In our experiments with DGS we found that there was not clear advantage to adopt either approach. Since greedy initial assignment takes a bit longer, we opted to use  random agent/job assignment for the initial solution. 
\subsection{Difference Evaluation}\label{ssec:difeval}  Starting from a full job/agent assignment $\sigma$, each agent tries to find the best 2-exchange in the neighborhood of $\sigma$. For each agent $i$ we consider how the objective function $f(\sigma)$ would change if it were to swap jobs with another agent $i'$ (i.e. a 2-exchange). We select the 2-exchange that yields the best improvement $\overline{\delta_i}$ and save it as agent $i$'s best configuration $NA_i$.  The procedure is called the \emph{agent difference evaluation} (ADE) and is described formally in Algorithm~\ref{alg:ade}.
Similarly, a \emph{job difference evaluation} (JDE) is carried out for each job, but in this case we consider swapping agents. 
\subsection{Switching}\label{ssec:switching} Here we select the 2-exchange that yields the greatest improvement in objective function value and modify the job/agent assignment accordingly. We then carry out JDE and ADE for the jobs and agents involved in that 2-exchange. We repeat the switching step until no further improvements are attainable. 

We define an assignment as a mapping $\sigma : J \to I$, where $J$ is the set of jobs and $I$ is the set of agents. Here $\sigma (j) = i$ means that job $j$ is assigned to agent $i$. Similarly another assignment mapping $\tau : I \to J$ is for mapping jobs to agent where $\tau(i) = j$ means that agent $i$ is assigned to job $j$. There is also an assignment mapping function to construct $\tau$ from $\sigma$ defined as $\tau = M(\sigma)$ and the objective function value of an assignment $\sigma$ is given by $f(\sigma)$. 
We make use of a switching function $\textbf{\textsc{switch}}
(i, j, \sigma)$ which returns a modified version the assignment $\sigma$ after agent $i$ has been assigned to job $j$; i.e. a 2-exchange has occurred between agents $i$ and $\sigma(j)$. For agent $i$, the job $j_i$ is the job that yields the largest increase in objective function when assigned to agent $i$ and it can be expressed as 
\begin{equation*}
 j_i=\argmax_{j=1,\ldots,n, j\neq \tau(i) }f(\textbf{\textsc{switch}}
 (i, j, \sigma))-f(\sigma),
\end{equation*}
and the corresponding improvement in objective function value is expressed as
\begin{equation*}
\overline{\delta}_i=\max_{j=1,\ldots,n, j\neq \tau(i) } f(\textbf{\textsc{switch}} (i, j, \sigma))-f(\sigma).
\end{equation*}
We similarly define for each job $j$ the best agent that it can be assigned to $ i_j$ and the corresponding improvement $\underline{\delta}_j$. Using this terminology, the algorithm is formally described in Algorithm \ref{alg:dgs}.

 \begin{algorithm}[t]\label{alg:dgs}
\SetArgSty{textit}
\DontPrintSemicolon
\SetKwComment{tcp}{$\triangleright\ $}
\textbf{\textsc{Algorithm DGS}} ($\sigma, f$)\\
{
	\Repeat{$f(\sigma_{start}) = f(\sigma')$}
	{
		$\sigma_{start} \gets \sigma$, $\tau=M(\sigma)$,
                $\overline{\delta} \gets \emptyset$, $\underline{\delta} \gets \emptyset$  \\
                $ADE(i,f,\tau,\sigma,NA,\overline{\delta})$ $\quad \forall i \in I$ \tcp*{Difference Evaluation}
                $JDE(j,f,\tau,\sigma,NJ,\underline{\delta})$ $\quad \forall j \in J$ \tcp*{Difference Evaluation}
		\While(\tcp*[f]{Switching phase} ){$\exists  \overline{\delta}_i > 0 \vee \exists  \underline{\delta}_j > 0 $}            
		{
			$i^* \gets \argmax_{i=1\ldots n} \overline{\delta}_i $, $j^* \gets \argmax_{j=1\ldots n} \underline{\delta}_j $ \\
			\eIf {$\overline{\delta}_{i^*}  > \underline{\delta}_{j^*}$}  {
			    $\overline{\delta}_{i^*} =0$\\
			    $\sigma' \gets \textbf{\textsc{switch}} (i^*, j_{i^*}, \sigma)$, $\tau' \gets M(\sigma')$ \\
			    $agents \gets \{i^*, \sigma'(\tau(i^*)) \}$, $jobs \gets \{\tau(i^*), \tau'(i^*) \}$
			}
			{
			    $\underline{\delta}_{j^*}=0$\\
			    $\sigma' \gets \textbf{\textsc{switch}} (i_{j^*}, j^*, \sigma)$\\
			    $agents \gets \{\sigma(j^*), \sigma'(j^*)) \}$, $jobs \gets \{j^*, \tau(\sigma'(j^*)) \}$
			}
			\If {$f(\sigma') > f(\sigma)$}
			{
				$\sigma \gets \sigma'$, $\tau=M(\sigma)$\\
			 $ADE(j,f,\tau,\sigma,NA,\overline{\delta})$ $\quad \forall i \in agents$\\
                         $JDE(j,f,\tau,\sigma,NJ,\underline{\delta})$ $\quad \forall j \in jobs$ \\
			}
		}
	}
	output $\sigma'$
}
\caption{DGS} 
\end{algorithm}

 \begin{algorithm}[th]\label{alg:ade}
\SetArgSty{textit}
\DontPrintSemicolon
\SetKwComment{tcp}{$\triangleright\ $}
\textbf{\textsc{Algorithm ADE}} ($i, f, \tau, \sigma, NA, \overline{\delta}$)\\
{
	$j \gets \tau(i)$, $\sigma_{i}^* \gets \sigma$, $\overline{\delta_i} \gets 0$\\
	\ForEach{$j' \in \{J ~ | ~ j' \neq j\}$}
	{
		$i' \gets \sigma(j')$ \\
		  $\sigma'_{i} \gets \sigma$, $\sigma'_{i}(j) \gets  i'$,  $\sigma'_{i}(j') \gets i $\\
		  \If{$f(\sigma'_{i}) > f(\sigma_{i}^*)$}
			{
				$\sigma_{i}^* \gets \sigma'_{i}$ \\
			}
	}
	\eIf{$\sigma_{i}^* \neq \sigma$}
	{
		$NA_i \gets \sigma_{i}^*$\\
		$\overline{\delta_i} \gets f(\sigma_{i}^*) - f(\sigma)$ \\
	}
	{
	      $NA_i \gets 0$ \\
	}
        
} 
\caption{ADE}
\end{algorithm}
\section{Evaluation}\label{sec:eval}
While explaining the process of realization of the CUDA solver in the
next section, we
also show results of the impact of the various steps that we went through to
implement it and enhance its performance. The experimental
setup for the tests consists of a consumer machine
with a 2.4Ghz Core 2 Duo processor equipped with 4GB of DDR3 RAM
and NVIDIA GTX 295 graphic card with 1GB of DDR5 on-board
memory. The NVIDIA GTX 295 is currently NVIDIA's top-of-the-line consumer
video card and boasts a total number of 30 Scalar Multiprocessors and
240 Processors, 8 for each SM, which run at a clock rate of 1.24 GHz.
In the experiments, we use a thread block size $256$ when executing kernels which do
not make use of Shared Memory, and $16$ in the case they do. 
Concerning the DGS input scenarios, we use dense instances of the GEOM type
defined by Bus and Tvrd{\i}k~\cite{geom}, and generated as follows: first
we generate $n$ points randomly in a 2D square of dimensions $[0,C]\times[0,C]$, then each  $a_{ij}$ value is set as the Euclidean distance
between points $i$ and $j$ from the generated $n$ points.  We define
the problem size to be equal to the number of agents/jobs. For the sake of simplicity, we make use of
problem sizes which are multiples of the thread block size. 
Note that every experiment is the result of the averaging of a number of
runs executed using differently seeded DGS instances.

\section{The DGS CUDA Solver}\label{sec:dgscuda}
The first prototype of the DGS solver was implemented in the Java
language. However, its performance did not meet the demands of our 
target real-time system. We therefore ported the same algorithm to
the C language in the hope that we obtain better performance. The outcome
of this effort was the first production-quality implementation of the
DGS which was sufficiently fast up to problem sizes of $5000$ peers.
The Difference Evaluation step of the algorithm, as described in Section~\ref{ssec:difeval} amounted to as much as $70$\% of the total
computational time of the solver. 
Luckily, all JDE and ADE evalution for
jobs and agents can be done in parallel as they are completely
orthogonal and they do not need to be executed
in a sequential fashion. 
Hence, our first point of investigation was to implement a CUDA ADE/JDE
kernel which could execute both ADE and JDE algorithms on the
GPU.

\begin{figure}[t]
\centering
\includegraphics[scale=0.6]{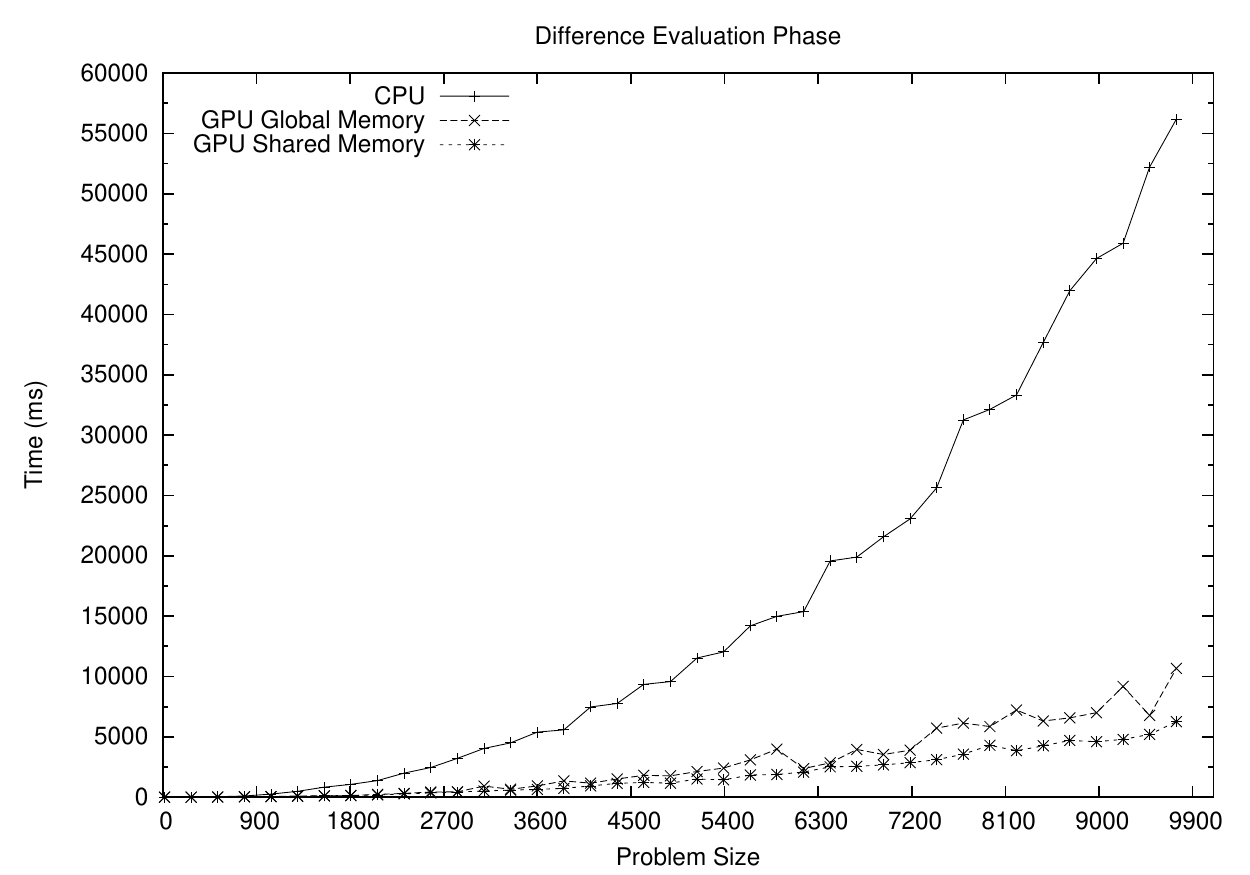}
\centering
\caption{Computational time comparison between Difference
  Evaluation implementations}
\label{fig:diff}
\end{figure}
We developed two versions of the ADE/JDE kernel: the first runs
exclusively on the GPU's Global memory and the second makes use of
the GPU's Shared memory to obtain better performance. For ease of exposition we will only 
discuss ADE going forward. This is without any loss of generality as everything that applies to ADE also applies to JDE, with the proviso the talk of jobs instead of agents.

\noindent \textbf{Difference Evaluation on Global Memory}\\
As mentioned earlier, Global memory is fully addressable by any thread
running on the GPU and no special operation is needed to access
data on it. Therefore, in the first version of the kernel, we decided
to simply
upload the full $A_{ij}$ matrix to the GPU memory together with the current
agent to job assignments and all the data we needed to run the ADE
algorithm on the GPU. Then we let the GPU spawn a thread for each
of the agents involved. Consequently, a CUDA thread $ct_i$ associated with
agent $i$ executes the ADE algorithm only for agent $i$ by
evaluating all its possible 2-exchanges. The  agent-to-thread allocation on 
the GPU is trivial and is made by assigning the thread identifier $ct_i$ to agent $i$. 

\noindent \textbf{Difference Evaluation on Shared Memory}\\
Difference Evaluation using
Shared Memory assigns one thread to each 2-exchange evaluation
for agent $i$ and job $j$. That implies that the number of created threads
equals the number of cells of the $A_{ij}$ matrix. 
 Each thread $ct_{ij}$ then proceeds to load in shared memory the data
which is needed for the single evaluation between agent $i$ and job
$j$. 
Once the
2-exchange evaluation is computed, every thread $ct_{ij}$  stores the resulting value in a
matrix located in global memory in position ($i$,$j$). After that, another small kernel is
executed which causes a thread for each
row $i$ of the resulting matrix to find the best 2-exchange value along that same
row for all indexes $j$. The outcome of this operation represents the best 2-exchange value for agent $i$. 
In Figure~\ref{fig:diff}, we compare the results obtained by running
the two aforementioned Shared Memory GPU kernel implementations and its
Global Memory counterpart against the pure C implementation of
the Difference Evaluation for different problem sizes. For evaluation purpose, we used a CUDA-enabled version of
the DGS where only the Difference Evaluation phase of the algorithm
runs on the GPU and can be evaluated separately from the other
phase. This implies that we need to upload the input data for the ADE/JDE phase to the GPU at every iteration of
the DGS algorithm, as well as we need to download its output in order to provide input for the Switching
phase. The aforementioned memory tranfers are accounted for in the
measurements. As we can observe, there's a dramatic improvement when
passing from the CPU implementation  of the difference evaluation to both GPU
implementations, even though multiple memory tranfers occur. In addition, the Shared Memory version behaves
consistently better than the Global Memory one. Furthermore, the trend for
increasing problem sizes is linear for both GPU versions of the
Difference Evaluation, opposed to the exponential growth of the CPU
version curve. 

\begin{algorithm}[t]\label{alg:dgsmod}
\SetArgSty{textit}
\DontPrintSemicolon
\SetKwComment{tcp}{$\triangleright\ $}
\textbf{\textsc{Algorithm DGS}} ($\sigma, f$)\\
{
	\Repeat{$f(\sigma_{start}) = f(\sigma')$}
	{
		$\sigma_{start} \gets \sigma$, $\tau=M(\sigma)$,
                $\overline{\delta} \gets \emptyset$,
                $\underline{\delta} \gets \emptyset$ \\
		\textbf{start parallel}  $\forall i \in I$,  $\forall j
                 \in J$     \tcp*{Difference Evaluation Phase starts}
                $ADE(i,f,\tau,\sigma,NA,\overline{\delta})$ \\
                $JDE(j,f,\tau,\sigma,NJ,\underline{\delta})$ \\
		\textbf{stop parallel}   \tcp*{Difference Evaluation Phase ends } 
		\While(\tcp*[f]{Switching phase} ){$\exists  \overline{\delta}_i > 0 \vee \exists
                  \underline{\delta}_j > 0 $ }  
		{
                $\textit{CC}(I,J,NA,NJ,C)$ 
		$\delta_i \gets 0 \quad \forall i \in I$, $\delta_{n+j} \gets 0 \quad \forall j \in J$  \\
                $\delta_i \gets \{ \overline{\delta}_i ~ | ~ i \notin C\} \quad \forall i \in I$ \\
                $\delta_{n+j} \gets \{ \underline{\delta}_j ~ |~ \sigma(j) \notin C\} \quad \forall j \in J$ \\
                \bf{start parallel} $\forall \delta_t>0$ \\
		\eIf {$t \leq n$}
		{
 			$i \gets t$, 
			$\sigma' \gets \textbf{\textsc{switch}} (i, j_{i}, \sigma)$ \\
		}
		{
			$j \gets (t-n)$, $\sigma' \gets \textbf{\textsc{switch}} (i_{j}, j, \sigma)$\\
		}			    
  		\If {$f(\sigma') > f(\sigma)$}
		{
			$\sigma \gets \sigma'$, $\tau=M(\sigma)$\\			 
		}
		\bf{stop parallel} \\
                \bf{start parallel} $\forall i \in  \{I ~| ~ i \notin
                C \}, \forall j \in \{J ~| ~ \sigma(j) \notin C \}$ \\
                   $ADE(i,f,\tau,\sigma,NA,\overline{\delta})$\\
                   $JDE(j,f,\tau,\sigma,NJ,\underline{\delta})$ \\
		\bf{stop parallel} \\
                 } 
	}
	output $\sigma'$
}
\caption{Parallel DGS} 
\end{algorithm}

 
 \begin{algorithm}[t]\label{alg:crc}
\SetArgSty{textit}
\DontPrintSemicolon
\SetKwComment{tcp}{$\triangleright\ $}
\textbf{\textsc{Algorithm CC}} ($I,J,NA,NJ,C$)\\
{
       $CR \gets \emptyset$, $C
                \gets \emptyset$\\
       \ForEach{$i \in \{I ~ | ~ NA_i \neq 0 \}$}
	{
                $\sigma \gets NA_i$\\
                $i' \gets \sigma(j_i)$\\
                \eIf{$i \in CR$ {\bf or} $i' \in CR$}
                {
                    $C \gets \{C, i\} $
                }
                {
                     $CR \gets \{CR,i,i'\}$
                }
        }
        \ForEach{$j \in \{J ~ | ~ NJ_j \neq 0 \}$}
	{
                $\sigma \gets NJ_j$ \\
                $i \gets \sigma(j)$\\
                \eIf{$i \in CR$ {\bf or} $i_j \in CR$}
                {
                    $C \gets \{C,i\} $
                }
                {
                     $CR \gets \{CR,i,i_j\}$
                }
        }   
} 
\caption{Check Conflicts}
\end{algorithm}


\noindent \textbf{Switching}\\
Considering the Switching phase of the DGS algorithm described in
Subsection \ref{ssec:switching} we found out that in
many cases the computational time necessary to apply the best
2-exchanges is fairly high. Our experience is that the switching phase
might have a relative impact of between 35\% and
60\% of the total computation time of the solver.
In order to improve the performance of this phase, we
modified the Switching algorithm so that a subset of the best 2-exchanges
computed in the Difference Evaluation section might be applied concurrently. 
The modified DGS algorithm is shown in Algorithm~\ref{alg:dgsmod}. In
order to execute some of the switches in parallel, we need to identify
which among them are not conflicting. For that, we designed a
function called \emph{CC}, shown in Algorithm~\ref{alg:crc}, which
serves the aforementioned purpose. Once the non-conflicted 2-exchanges are determined by CC,
we identify the corresponding agents and jobs and we apply the
exchanges in a parallel fashion.
After this operation completes, we proceed to re-evaluate the differences for
the agents and jobs whose 2-exchanges were identified as conflicting,
for there might be possible better improvements for those which were
not applied. At the next iteration of the DGS algorithm, conflicted
two-exchanges may be resolved and applied in parallel. In order to
execute the parallel Switching phase on the GPU, we simply let the GPU spawn a number
of threads which is equal to the number of non-conflicting
2-exchanges and let them perform the switch. 
\section{Results}\label{sec:meas}
\begin{figure*}[t]
\centering
\includegraphics[scale=0.7]{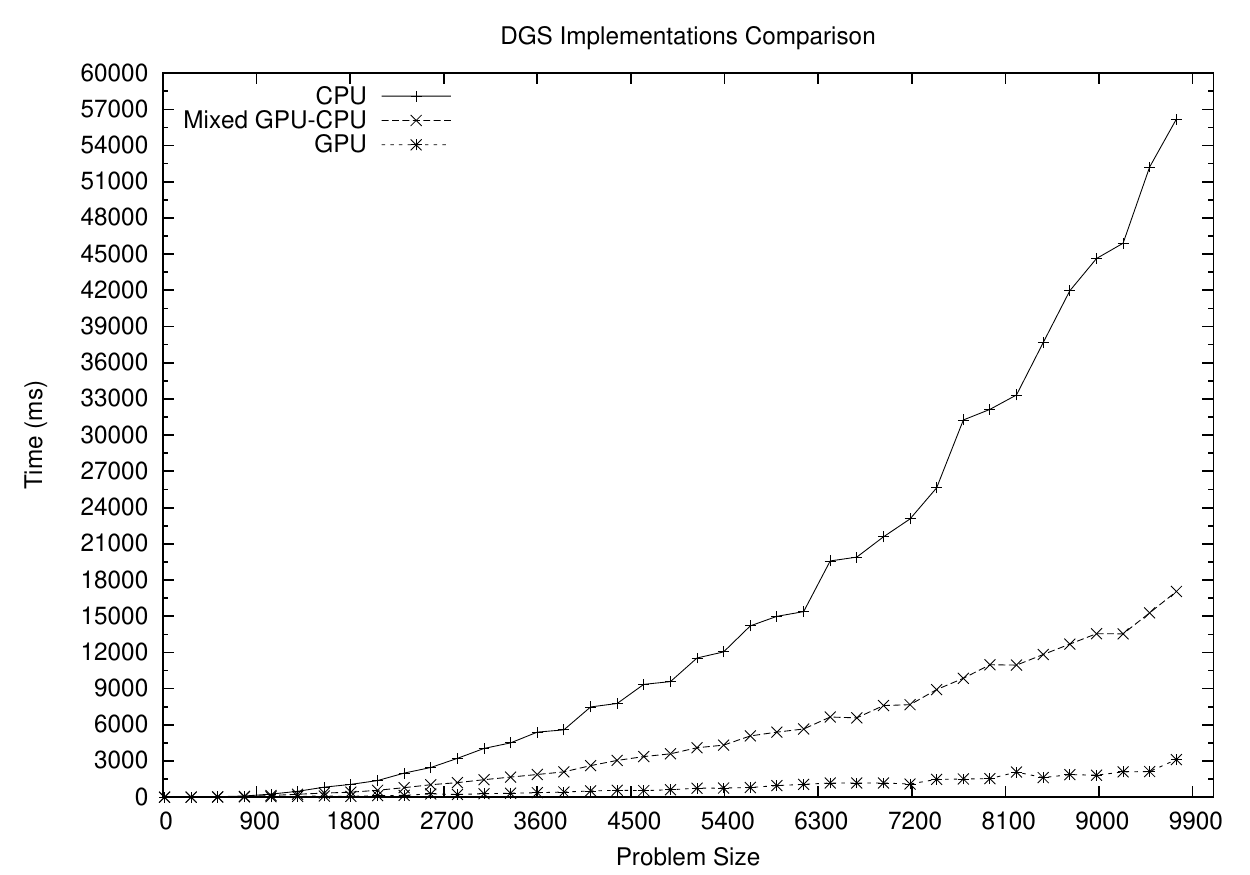}
\centering
\caption{Computational time comparison between DGS's implementations.}
\label{fig:dgs}
\end{figure*}

In Figure~\ref{fig:dgs} we show the results obtained by comparing three
different implementations of the DGS heuristic: a pure C
implementation labelled ``CPU'', the ``Mixed GPU-CPU'' implementation, where only the
Difference Evaluation section of the algorithm is
executed on the GPU using Shared Memory, and the ``GPU''
implementation, where all three main phases of the
DGS including the Switching are executed on the GPU. As we can
observe, the gain in performance when considering the ``GPU''
compared to the two other implementations is paramount. There are two
fundamental reasons for that. The first is the speed-up obtained by
applying all non-conflicting 2-exchanges in parallel. 
The second reason
 is a direct consequence of the fact that most of the operations are
executed directly on the GPU and few host--device operations are
needed. Such operations,
e.g. memory transfers,
can be expensive and certainly contribute to the absolute time needed for the solver
to reach an outcome. In fact, it's interesting to observe that the
total termination time needed for big problem sizes is less than the total time
needed for executing just the ADE/JDE phase, as shown in Figure~\ref{fig:diff}, where multiple memory tranfers occur at every
iteration of the algorithm.  
\begin{figure*}[t]
\centering
\includegraphics[scale=0.7]{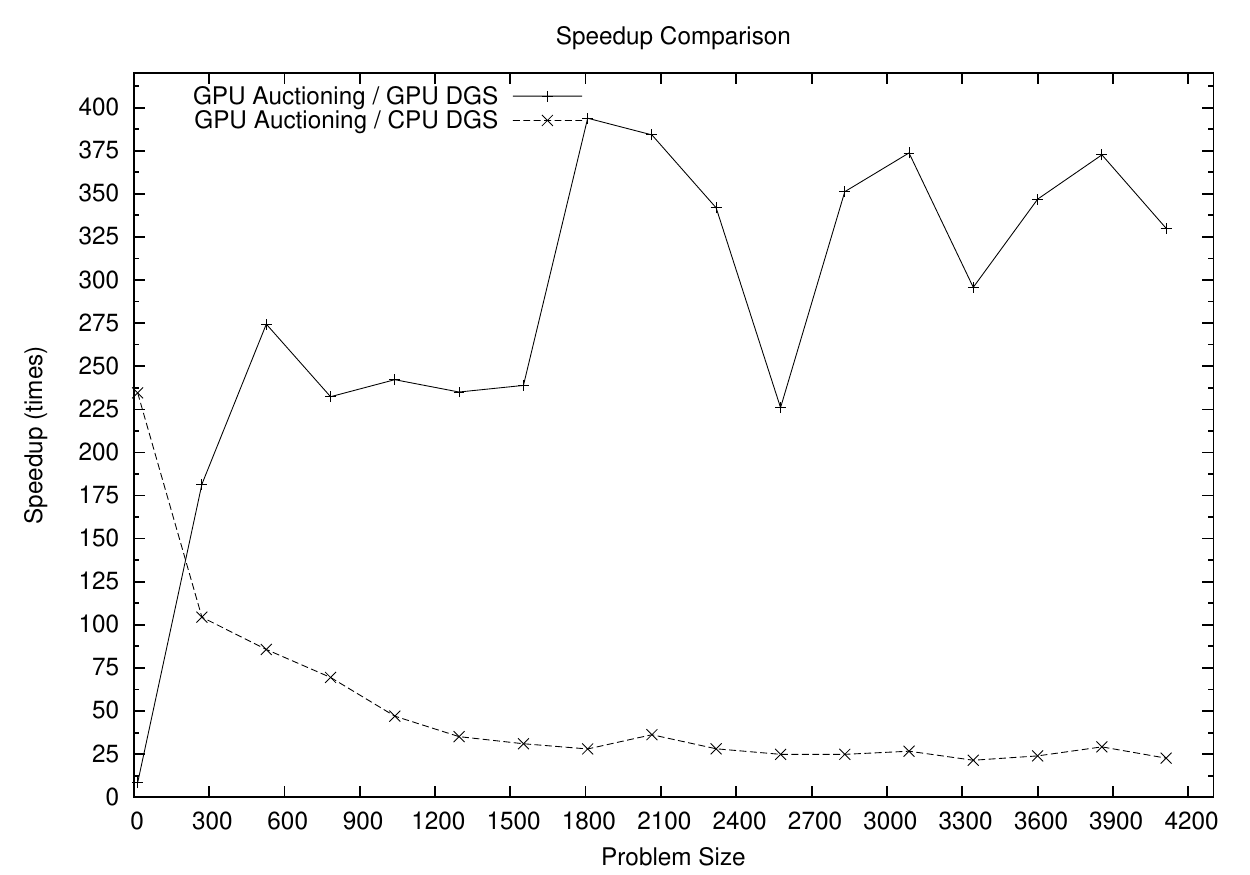}
\centering
\caption{Speed-up comparison between DGS's implementations compared to GPU Auctioning}
\label{fig:speedup}
\end{figure*}
In order to assess the improvement in performance of our GPU DGS solver with respect
to other LSAP solvers,  we compare our solver to an
implementation of the Auction algorithm 
published by Vasconcelos et al.~\cite{gpuauction}, which is implemented on GPUs using
the CUDA language. Figure~\ref{fig:speedup} shows the outcome of this analysis. As
we can observe, the speed--up obtained can be as high as 400 times faster. Furthermore,
we can note that even the CPU version of the sequential DGS algorithm performs considerably better
than the GPU auctioning solver, as much as 20 times faster for large problem sizes.
\section{Conclusion \& Future Work}\label{sec:conclusion}
In this paper we presented the realization of a GPU-enabled LSAP solver
based on the Deep Greedy Switching heuristic algorithm and implemented
using the CUDA programming language. We detailed the process of
implementation and enhancement of the two main phases of the
algorithm: Difference Evaluation and Switching, and we
provided results showing the impact of each iteration on the
performance. In particular, we showed how parallelizing some parts of the
solver with CUDA can lead to substantial speed--ups. We also suggested a modification to
the DGS algorithm, in the Switching phase, which enables
the solver to run entirely on the GPU. In the last part of the paper, we also show the
performance of the final version of the solver compared to a pure C
language DGS implementation and to an
auction algorithm implementation on GPUs, concluding that
the time needed for the DGS solver to reach an outcome is one order
of magnitude lower compared to the ``C'' implementation for big scenarios
and three orders of magnitude lower on average compared to the
GPU auction solver in almost all problem sizes. 
For future work, we would like to formally analyze the modified
version of  the DGS algorithm to theoretically assess its lower bound
on optimality. We would also like to see our solver applied in different 
contexts and explore possible applications involving LSAP that have yet to be investigated due to computational limitations. 

\bibliographystyle{IEEEtran}
\bibliography{GPUDGS}

\end{document}